\documentclass[review]{elsarticle}

\usepackage{lineno,hyperref}
\modulolinenumbers[5]

\usepackage{multirow,eurosym,amssymb,amsfonts,amsmath,setspace,graphicx,color,bm,float,verbatim,subfigure}

\newcommand{\ket}[1]{\left|#1\right\rangle}
\newcommand{\bra}[1]{\left\langle #1\right|}

\newcommand{\blue}{\color[rgb]{0.0,0.0,0.6}}
\def\ket#1{| #1 \rangle}
\def\bra#1{\langle #1 |}

\def\be{\begin{equation}}
\def\ee{\end{equation}}

\def\bsplit{\begin{split}}
\def\nsplit{\end{split}}
\journal{Optics Communications}

\begin{document}

\begin{frontmatter}

\title{Generating non-classical states from spin coherent states via interaction with ancillary spins}

\author{Shane Dooley }
\address{Quantum Information Science, School of Physics and Astronomy, University of Leeds, Leeds LS2 9JT, U.K.}

\author{Jaewoo Joo \corref{mycorrespondingauthor}}
\ead{j.joo@leeds.ac.uk; +441133435036}
\address{Quantum Information Science, School of Physics and Astronomy, University of Leeds, Leeds LS2 9JT, U.K.}

\author{Timothy Proctor}
\address{Quantum Information Science, School of Physics and Astronomy, University of Leeds, Leeds LS2 9JT, U.K.}

\author[mymainaddress,mysecondaryaddress]{\rm Timothy P. Spiller \corref{mycorrespondingauthor}}
\cortext[mycorrespondingauthor]{Corresponding author}
\address[mymainaddress]{Quantum Information Science, School of Physics and Astronomy, University of Leeds, Leeds LS2 9JT, U.K.}
\address[mysecondaryaddress]{York Centre for Quantum Technologies, Department of Physics, University of York, York YO10 5DD, U.K.}

\begin{abstract}
The generation of non-classical states of large quantum systems has attracted much interest from a foundational perspective, but also because of the significant potential of such states in emerging quantum technologies. In this paper we consider the possibility of generating non-classical states of a system of spins by interaction with an ancillary system, starting from an easily prepared initial state . We extend previous results for an ancillary system comprising a single spin to bigger ancillary systems and the interaction strength is enhanced by a factor of the number of ancillary spins. Depending on initial conditions, we find -- by a combination of approximation and numerics -- that the system of spins can evolve to spin cat states, spin squeezed states or to multiple cat states. We also discuss some candidate systems for implementation of the Hamiltonian necessary to generate these non-classical states.
\end{abstract}


\end{frontmatter}

\section{Introduction}

The apparent conflict between the classical world of our everyday experience and the underlying quantum reality has been widely discussed since the beginnings of quantum physics. A prominent line of research is the effort to create entanglement between a macroscopic quantum system and a microscopic system along the lines of the original Schr\"{o}dinger cat thought experiment \cite{Schrodinger,Wheeler83}. The generation of macroscopic quantum states is also interesting from a technological point of view. In optical systems, for example, it is known that various macroscopic non-classical states can be exploited to give a significant improvement in the precision of phase estimation \cite{Noon, Caves}. Along these lines, many proof-of-principle experiments have been demonstrated with optical systems \cite{Knight}. However, hybrid quantum systems might be eventually needed to extract quantum advantages in practical quantum technology because alternative physical set-ups could provide different advantages. % (e.g., intialisation in spin states).

Recently, continuous-variable (CV) superposed/entangled states have shown their potential in various optical and photonic experiments \cite{Jeong13} and the use of CV entangled states can be robust in practical quantum metrology \cite{ECS1,ECS2,ECS3}. Recently, ``micro-macro'' entangled states have been implemented from a path-entangled single photon state  \cite{Gisin13,Lvo-13}. The beam splitting interaction, by putting a vacuum in one mode and a single photon Fock state in the other, followed by a displacement in one of the modes leads to the micro-macro entangled state $D_B(\alpha) (\ket{1}_A \ket{0}_B + \ket{0}_A \ket{1}_B)/\sqrt{2}$ where $D_B(\alpha)$ is a displacement operator with amplitude $\alpha$ in mode $B$ \cite{Lvo-13}. % and the macroscopic superposed state could be obtained in mode $B$ at large $\alpha$ by measuring the photon in mode $A$. 

Here, we consider the generation of non-classical states of two interacting systems $A$ and $B$ where system $A$ is a collection of $N_A$ spin-$1/2$ particles and ancillary system $B$ is a collection of $N_B$ spins. Spin states have previously been considered as a way of storing a qubit (i.e., two orthogonal collective spin states are used as computational basis states of an effective qubit \cite{Polzik10,Polzik99,Polzik14,NV-Lukin13}) but can be naturally utilised for CV quantum information processing by creating CV entangled states in a spin system. We assume that the initial state consists of each of the $N_A$ qubits in the same pure state, an easily prepared state in principle. In \cite{Shane13} and \cite{Doo-14} it was shown that various CV states (e.g., spin cat states, multiple cat states, and spin squeezed states) can be generated from a spin coherent state (SCS) for $N_B=1$. By a combination of approximation and numerics, we investigate cases when $N_B > 1$. An advantage of $N_B > 1$ compared to $N_B=1$ is faster preparation times of the non-classical CV states. 

This paper is organised as follows. In Section \ref{Sec_2}, we give the interaction Hamiltonian and show that it has several well-known Hamiltonians as limits. In Section \ref{Sec_3} we present short-time approximations for the dynamics of the model for two different initial states of spin system $A$. In the first case, we show that for a carefully chosen initial state of the ancillary system $B$, the system $A$ evolves to a superposition of two spin coherent states, a spin ``Schrodinger cat'' state. In the second approximation we show that spin system $A$ evolves to spin squeezed states. We also give numerical evidence that at later times, beyond the restriction of the approximation, the combined system $AB$ can evolve to a superposition of many spin coherent states (``multiple cat states'') of the combined system. We suggest an ansatz Hamiltonian that predicts the gross features of the dynamics in this case.

In Section \ref{Sec_4}, we discuss a beam-splitter (BS) type interaction between two distinct systems of spins. If a non-classical SCS interacts with a typical SCS, the resultant state can be understood as an entangled CV SCS in two modes. This has potential to be implemented in Bose-Einstein condensates (BECs) and Nitrogen-vacancy centres (NV-centres) with superconducting systems. Finally, we summarise the results in Section \ref{Sec_5}.

\section{Spin Hamiltonian model}
\label{Sec_2}
Let us assume that system $A$ is a collection of $N_A$ spins and system $B$ is that of $N_B$ spins. We consider a Hamiltonian of the form
\begin{equation} 
\hat{H}(N_A, N_B) =  \omega_A \left(\hat{J}_A^{z} + \frac{N_A}{2} \right)  +   \omega_B \left( \hat{J}_B^{z} + \frac{N_B}{2} \right) +\lambda \left( \hat{J}_{A}^{+}\hat{J}_{B}^{-} + \hat{J}_{A}^{-} \hat{J}_{B}^{+}  \right) , \label{eq:HAB} 
\end{equation} 
where the $J$-operators on system $A$ are defined as 

\begin{equation} \hat{J}_A^{\mu} = \frac{1}{2} \sum_{i = 0}^{N_A} \hat{\sigma}_{(i)}^{\mu} \quad ; \quad \hat{J}_A^{\pm} =  \sum_{i = 0}^{N_A} \hat{\sigma}_{(i)}^{\pm} \quad ; \quad \hat{J}_A^2 =  \sum_{\mu} (\hat{J}_A^{\mu})^2 , \end{equation} where $\hat{\sigma}^{\mu}$ are the Pauli operators for the individual spins of $A$ with $\mu \in \{x,y,z\}$. The $J$-operators for $B$ are defined in the same way. 

The \emph{Dicke states} are the set of simultaneous eigenstates of the commuting operators $\hat{J}_A^2$ and $\hat{J}_A^z$, and are denoted by $\ket{j,n-j}_{A}$ where \begin{equation} \hat{J}_A^2 \ket{j,n-j}_{A} = j(j+1)\ket{j,n-j}_{A} \quad ; \quad \hat{J}_A^z \ket{j,n-j}_{A} = (n-j)\ket{j,n-j}_{A} \end{equation} for $j\in\left\{0,1,...,\frac{N_A}{2} \right\}$ if $N_A$ is even, $j\in\left\{\frac{1}{2},\frac{3}{2},...,\frac{N_A}{2} \right\}$ if $N_A$ is odd, and $n\in\left\{0,1,...,2j \right\}$. States in the $j=\frac{N_A}{2}$ eigenspace of the $N_A$ spin system are totally symmetric with respect to exchange of any two spins. In particular, the $j=\frac{N_A}{2}$ Dicke states are totally symmetric:

\begin{equation} \ket{\frac{N_A}{2}, n- \frac{N_A}{2}}_A = \binom{N_A}{n}^{-1/2} \sum_{\mbox{permutations}} \ket{\downarrow^{\otimes (N_A -n)} \uparrow^{\otimes n}} , \label{eq:dicke}\end{equation} where $\ket{\uparrow}$ and $\ket{\downarrow}$ are eigenstates of $\sigma^z$ for a single spin. The $N_A + 1$ Dicke states ($n\in\left\{ 0,1,...,N_A \right\} $) are a basis for the $j=\frac{N_A}{2}$ eigenspace (this is true only for this eigenspace, the one associated with the maximal value of $j$). In what follows we restrict to the $j=\frac{N_A}{2}$ eigenspace and the Dicke state in Eq.~(\ref{eq:dicke}) is written as $\ket{\frac{N_A}{2}, n- \frac{N_A}{2}}_A \equiv \ket{n}_A$ for simplicity. %It is convenient to restrict to this subspace because...

The $\hat{J}_A^{\pm}$ operators have the effect of raising and lowering the $n$ index of Dicke states:

\begin{eqnarray} \hat{J}_A^{+}\ket{n}_{A} &=& \sqrt{\left( n+1 \right)\left( N_A -n \right)} \ket{n+1}_{A} , \\ \hat{J}_A^{-}\ket{n}_{A} &=& \sqrt{n\left( N_A -n+1 \right)} \ket{n-1}_{A} . \end{eqnarray} 

These raising and lowering operators have a superficial similarity to the creation and annihilation operators of a bosonic field mode, \begin{equation}  \hat{a}^{\dagger}\ket{\bar{n}} = \sqrt{n+1}\,\ket{\overline{n+1}} \quad ; \quad  \hat{a}\ket{\bar{n}} = \sqrt{n}\,\ket{\overline{n-1}} , \end{equation} where $\ket{\bar{n}}$ are Fock states, eigenstates of $\hat{a}^{\dagger}\hat{a}$ (the bar above $n$ indicates a state of the field mode rather than a state of the finite spin system). In fact, it is not difficult to see that if we identify the Dicke state $\lim_{N_A \to\infty}\ket{n}_A$ with the Fock state $\ket{\bar{n}}$ of a bosonic mode, then

\begin{equation} \lim_{N_A \to\infty} \frac{\hat{J}_A^+}{\sqrt{N_A}} = \hat{a}^{\dagger} \quad ; \quad \lim_{N_A \to\infty} \frac{\hat{J}_A^-}{\sqrt{N_A}} = \hat{a} , \end{equation} and $\hat{J}_A^{\pm}/\sqrt{N_A}$ obey the bosonic commutation relations: 

\begin{equation} \lim_{N_A \to\infty} \left[ \frac{\hat{J}_A^{-}}{\sqrt{N_A}}, \frac{\hat{J}_A^{+}}{\sqrt{N_A}} \right] = 1  . \end{equation} This is the \emph{bosonic limit} of the spin raising and lowering operators. If $N_A$ is finite then we have

\begin{equation} \left[ \frac{\hat{J}_A^{-}}{\sqrt{N_A}}, \frac{\hat{J}_A^{+}}{\sqrt{N_A}} \right] = 1 - \frac{2}{N_A}\left( \hat{J}_A^z + \frac{N_A}{2} \right) , 
\label{eq:apprcom} 
\end{equation} and the spin raising and lowering operators approximately satisfy the bosonic commutation relations only if the second term on the right hand side of Eq.~(\ref{eq:apprcom}) can be neglected. For $N_A$ finite we also have the Holstein-Primikoff transformations  \cite{Hol-40} that relate the $J$-operators to the bosonic operators:

\begin{equation} \frac{\hat{J}_A^{-}}{\sqrt{N_A}} = \sqrt{1 - \frac{\hat{a}^{\dagger}\hat{a}}{N_A}} \; \hat{a} \quad ; \quad \frac{\hat{J}_A^{+}}{\sqrt{N_A}} =  \hat{a}^{\dagger} \sqrt{1 - \frac{\hat{a}^{\dagger}\hat{a}}{N_A}} \quad ; \quad \hat{J}_A^z = \hat{a}^{\dagger}\hat{a} - \frac{N_A}{2} .  \label{eq:HP}  \end{equation} As in Eq.~(\ref{eq:apprcom}), if the $\hat{a}^{\dagger}\hat{a}/N_A$ contributions under the square roots in Eq.~(\ref{eq:HP}) can be neglected, the $N_A$ spin system (in the $j=N_A/2$ subspace) is well approximated as a bosonic mode. 

The model Hamiltonian (\ref{eq:HAB}) has a number of other interesting models as special limits. To see this it is first useful to renormalise the interaction parameter to $\lambda = \tilde{\lambda}/\sqrt{N_A N_B}$ so that we get sensible results after taking limits. Then, for example, if we take the $N_A \to \infty$ limit and choose $N_B = 1$ we are left with the familiar Jaynes-Cummings Hamiltonian for the interaction of a bosonic mode with a two level system:

\begin{equation}  \hat{H}(\infty,1) = \omega_A \, \hat{a}^{\dagger}_A \hat{a}_A + \frac{\omega_B}{2}\left( \hat{\sigma}_{B}^z + 1 \right) + \tilde{\lambda}\left( \hat{a}_A \hat{\sigma}_B^{+}  +  \hat{a}^{\dagger}_A \hat{\sigma}_B^{-}  \right) . \end{equation} If we let $N_A \to \infty$ and allow $N_B$ to be some finite number we have the Tavis-Cummings Hamiltonian:

\begin{equation}  \hat{H}(\infty, N_B) = \omega_A \, \hat{a}^{\dagger}_A \hat{a}_A + \omega_B \left( \hat{J}_B^{z} + \frac{N_B}{2} \right) + \frac{\tilde{\lambda}}{\sqrt{N_B}}\left( \hat{a}_A \hat{J}_B^{+}  +  \hat{a}^{\dagger}_A \hat{J}_B^{-}  \right) . \end{equation} If we take both $N_A \to \infty$ and $N_B \to \infty$ we get

\begin{equation}  \hat{H}(\infty, \infty) = \omega_A \, \hat{a}^{\dagger}_A \hat{a}_A + \omega_B  \, \hat{b}^{\dagger} \hat{b}_B  + \tilde{\lambda}\left( \hat{a}_A \hat{b}^{\dagger}_B  +  \hat{a}^{\dagger}_A \hat{b}_B \right) , \end{equation} the Hamiltonian for an exchange interaction between two bosonic modes.

Each of these interaction Hamiltonians can be used -- in principle -- to generate macroscopic superposition states of, say, system $A$, and/or macroscopic entangled states of $AB$, starting from easily prepared initial states of $A$. The on-resonance Jaynes-Cummings model, for instance, with an initial mesoscopic coherent state and an appropriately chosen initial qubit state, evolves to a Schr\"odinger cat state of the field mode at a quarter of the revival time (via an entangled state of the field and the atom) \cite{Gea-91}. The on-resonance Tavis-Cummings Hamiltonian can be applied to generate the same Schr\"odinger cat state with a shorter evolution time, but with the cost that the $N_B$ qubits must be initially in a GHZ-type state \cite{TPS09}. 

Transforming the Hamiltonian (\ref{eq:HAB}) to the interaction picture with respect to the free Hamiltonian $\hat{H}_0 = \omega_B \left( \hat{J}_A^z + \hat{J}_B^z + \frac{N_A + N_B}{2} \right)$ gives the interaction picture Hamiltonian

\begin{equation} \hat{H}_I (N_A, N_B) = \Delta \left( \hat{J}_A^z + \frac{N_A}{2} \right) +\lambda \left( \hat{J}_{A}^{+}\hat{J}_{B}^{-} + \hat{J}_{A}^{-} \hat{J}_{B}^{+}  \right) , \label{eq:Hint} \end{equation} where $\Delta = \omega_A - \omega_B$ is the detuning. On resonance ($\omega_A = \omega_B$) this reduces to \begin{equation} \hat{H}_I (N_A, N_B) = \lambda \left( \hat{J}_{A}^{+}\hat{J}_{B}^{-} + \hat{J}_{A}^{-} \hat{J}_{B}^{+}  \right) . \label{eq:HABint} \end{equation} This interaction term allows for a collective, coherent excitation to be exchanged between system $A$ and system $B$.

Here, assuming resonance and assuming that both $N_A$ and $N_B$ are finite with $N_A \gg N_B$, we find interesting states generated by the model Hamiltonian in the parameter regime in which the bosonic approximation is applicable, as well as interesting states outside of this parameter regime. 

\section{Generation of non-classical states}
\label{Sec_3}
\subsection{Spin coherent states}
We assume that system $A$ is initially in a SCS \cite{Are-72} of the $N_A$ spins given

\begin{equation} \ket{\zeta}_A = \bigotimes_{i=1}^{N_A} \left[  \frac{ \ket{\downarrow^{(i)}} + \zeta \ket{\uparrow^{(i)}} }{\sqrt{1 + |\zeta|^2}} \right]   , \end{equation} 
where $\zeta$ is a complex number. This state is, in principle, easily prepared since each of the spins is in the same pure state.
Expanding the tensor product we can write this state in the Dicke basis as

\begin{equation} \ket{\zeta}_A =   \sum_{n=0}^{N_A} C_n \ket{n}_A , \label{eq:SCSdicke} \end{equation} where $C_n = \sqrt{\binom{N_A}{n}} \frac{\zeta^n}{ \sqrt{1+|\zeta|^2 }^{N_A}}$. Written in this form, it can be shown \cite{Mar-03,Shane13} that in the $N_A \to\infty$ limit the SCS $\ket{\frac{\zeta}{\sqrt{N_A}}}_A$ is identical to the bosonic coherent state $\ket{\bar{\zeta}}$ with complex amplitude $\zeta$:

\begin{figure}[h]
\includegraphics[width=10.3cm]{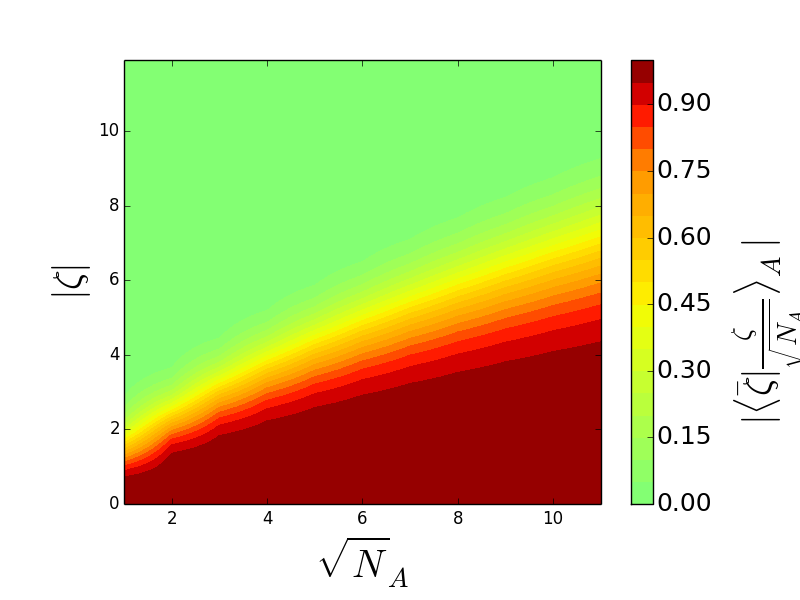}
\caption{The fidelity $|\langle\bar{\zeta}|\frac{\zeta}{\sqrt{N_A}}\rangle_A |$ is close to unity when $|\zeta|\ll \sqrt{N_A}$. The spin coherent state mimics a bosonic coherent state in this parameter regime.} \label{fig:fidCStoSCS}
\end{figure}

\begin{equation}  
\lim_{N_A \to\infty}\ket{\frac{\zeta}{\sqrt{N_A}}}_A = \ket{\bar{\zeta}} = e^{-|\zeta|^2/2} \sum_{n=0}^{\infty} \frac{\zeta^n}{\sqrt{n!}} \ket{\bar{n}} , \end{equation} where, again, we have identified the Dicke state $\lim_{N_A \to\infty}\ket{n}_A$ with the Fock state $\ket{\bar{n}}$. When $N_A$ is finite the SCS $\ket{\frac{\zeta}{\sqrt{N_A}}}_A$ is well approximated by the coherent state $\ket{\bar{\zeta}}$ when  $|\zeta| \ll \sqrt{N_A}$. In figure ~\ref{fig:fidCStoSCS}, the fidelity $|\langle\bar{\zeta}|\frac{\zeta}{\sqrt{N_A}}\rangle_A |$ is shown between a coherent state and a SCS  with respect to different values of $\zeta$ and the number of spins $N_A$. Thus, big spins are capable of mimicking an optical coherent state. % with large amplitude $\zeta$. 

\subsection{Approximate dynamics}

In this subsection (following the methods of \cite{Kli-95}) we approximate the on-resonance interaction picture Hamiltonian (\ref{eq:HABint}) with a big spin in mode $A$ and small fixed number of spins in mode $B$ ($N_A \gg N_B$), for the initial separable state 
\begin{equation} 
\ket{\psi_m (0)} = \ket{\zeta}_{A} \ket{D_m^{\phi}}_B ,
\end{equation}  where $\ket{\zeta}_{A}$ is a SCS in mode $A$ with 
\begin{equation} 
\sqrt{\frac{N_B }{ N_A}} \ll |\zeta| \ll {\sqrt{ \frac{N_A }{ N_B} }} , \label{eq:par}
\end{equation} 
and $\ket{D_m^{\phi}}_B = e^{-i\phi \hat{J}_B^z} \ket{D_m^0}_B$ is a Dicke state of $B$, a simultaneous eigenstate of $\hat{J}_B^2$ and the operator $\hat{J}_B^{+} e^{-i\phi} + \hat{J}_B^{-} e^{i\phi}$ with eigenvalue $m$. We note that $\ket{D_m^{\phi}}_B$ depends on the phase $e^{-i\phi} = \zeta/|\zeta|$ of the initial coherent state of system $A$.  These states are sometimes called ``semi-classical eigenstates'' \cite{Gea-91} of the Hamiltonian in Eq.~(\ref{eq:HAB}) because they are eigenstates of the Hamiltonian if we replace $\hat{J}_A^{+}$ and $\hat{J}_A^{-}$ with their expectation values in the initial SCS of $A$. 

We rewrite the Hamiltonian in terms of the operator $\hat{P}\equiv \left( \hat{J}_A^- \hat{J}_A^+ \right)^{-1/2} \hat{J}_A^-$:
\begin{equation}  \hat{H}_I (N_A, N_B) = \lambda \left( \sqrt{\hat{J}_A^{-} \hat{J}_A^{+} } \hat{P} \hat{J}_B^{+} + \hat{J}_B^{-} \hat{P}^{\dagger} \sqrt{\hat{J}_A^{-} \hat{J}_A^{+} }   \right) . \end{equation} We also define the operator

\begin{equation} \hat{Q} \equiv \hat{P}^{J_B^z +c} = \sum_{m_z = -\frac{N_B}{2}}^{N_B /2} \hat{P}^{m_z + c} \ket{m_z } \bra{m_z}  , \end{equation} where $c=0$ if $N_B$ is even and $c=\frac{1}{2}$ if $N_B$ is odd and $\ket{m_z}$ are the eigenstates of $J_B^z$. Neither the operators $\hat{P}$ or $\hat{Q}$ are exactly unitary, but are approximately unitary in the sense that they are unitary if they act on a subspace of $A$ that excludes the Dicke states $\ket{n}_A$ with $0\leq n \leq \frac{N_B}{2}+c$ and $N_A - \frac{N_B}{2}-c \leq n \leq N_A$. Restriction to this subspace is a good approximation for the initial spin coherent state, since the amplitudes corresponding to these Dicke states are negligible.

We now transform the frame of reference by the ``unitary'' operator $\hat{Q}$ so that the initial state is given by $\hat{Q}^{-1}\ket{\psi_m (0)}$ and the Hamiltonian becomes

\begin{eqnarray}  \hat{H}_I^{Q}  = \hat{Q}^{-1} \hat{H}_I \hat{Q} = \lambda \left( \hat{Q}^{-1} \sqrt{\hat{J}_A^{-} \hat{J}_A^{+} } \, \hat{Q} \; \hat{J}_B^{+} + \hat{J}_B^{-} \; \hat{Q}^{-1} \sqrt{\hat{J}_A^{-} \hat{J}_A^{+} } \, \hat{Q}  \right) , \label{eq:HQ}   \end{eqnarray} where we have used the identities: 

\begin{equation} \hat{Q} \hat{J}_B^{-} \hat{Q}^{-1} = \hat{P}^{\dagger} \hat{J}_B^{-} \quad ; \quad   \hat{Q} \hat{J}_B^{+} \hat{Q}^{-1} = \hat{P}  \hat{J}_B^{+} . \end{equation} 

We find that the operator $\hat{Q}^{-1} \sqrt{\hat{J}_A^{-} \hat{J}_A^{+} } \, \hat{Q}$ in (\ref{eq:HQ}) is given by \begin{equation}  \hat{Q}^{-1} \sqrt{\hat{J}_A^{-} \hat{J}_A^{+} } \, \hat{Q}  =  \sqrt{ \hat{J}_A^2 - (J_A^z -J_B^z -c)(J_A^z -J_B^z -c +1) } , \label{eq:end} \end{equation} assuming that it is acting on the subspace of system $A$ for which $\hat{Q}$ is unitary. 

The upshot of this transformation is that $\hat{P}$ has been eliminated from the Hamiltonian (\ref{eq:HQ}) which is now diagonal in the Dicke basis $\ket{n}_{A}$ since the operator $\hat{Q}^{-1} \sqrt{\hat{J}_A^{-} \hat{J}_A^{+} } \, \hat{Q}$ is diagonal (in the sense that ${}_{A}\bra{ n } \hat{Q}^{-1} \sqrt{\hat{J}_A^{-} \hat{J}_A^{+} } \, \hat{Q} \ket{n'}_{A} \propto \delta_{n n'}$). This is useful since the operator $\hat{a}^{\dagger} \hat{a} = \hat{J}_z + \frac{N_A}{2}$ (whose eigenstates are the Dicke states $\ket{n}_{A}$ with eigenvalue $n$) has expectation value and standard deviation 

\begin{equation} \langle \hat{a}^{\dagger} \hat{a} \rangle =  \frac{N_A |\zeta|^2}{1 + |\zeta|^2} \quad ; \quad \Delta ( \hat{a}^{\dagger} \hat{a}) = \frac{\sqrt{N_A} |\zeta|}{1 + |\zeta|^2} , \end{equation} for the initial state $Q^{-1}\ket{\psi_m (0)} = e^{i\phi c}\ket{\zeta}_{A} \ket{D_m^0}_B$ (in the transformed basis).

When $\sqrt{\frac{N_a }{ N_A}} \ll |\zeta| \ll {\sqrt{ \frac{N_A }{ N_a} }}$, we have that $\frac{\Delta ( \hat{a}^{\dagger} \hat{a})}{ \langle \hat{a}^{\dagger} \hat{a} \rangle} \ll 1$. This means that the distribution of the Dicke states $\ket{n}_{A}$ in the initial state $\hat{Q}^{-1}\ket{\psi_m (0)}$ is narrowly peaked around its average value $ \langle \hat{a}^{\dagger} \hat{a} \rangle$. This property is maintained as the state evolves since $\hat{H}_I^Q$ is diagonal in the Dicke basis $\ket{n}_{A}$. In the following paragraphs we use this property to make further approximations in two different parameter regimes, the first corresponding to the bosonic approximation when $\sqrt{N_B / N_A} \ll |\zeta| \ll 1$ , and the second in a very different parameter regime with $|\zeta| = 1$.

Firstly, we assume that the initial SCS parameter of system $A$ is in the range $\sqrt{N_B / N_A} \ll |\zeta| \ll 1$ so that $ \langle \hat{a}^{\dagger} \hat{a} \rangle \approx N_A |\zeta|^2 \gg N_a$ and $\Delta ( \hat{a}^{\dagger} \hat{a}) \approx \sqrt{\langle \hat{a}^{\dagger} \hat{a} \rangle} \gg \sqrt{N_B}$.
Since the distribution of the Dicke states $\ket{n}$ is always narrowly peaked around its average value, we may say that the terms that contribute significantly to the evolving state are those associated with the Dicke state label $n$ in the range $\langle \hat{a}^{\dagger} \hat{a} \rangle -  \Delta ( \hat{a}^{\dagger} \hat{a}) \leq n \leq \langle \hat{a}^{\dagger} \hat{a} \rangle + \Delta ( \hat{a}^{\dagger} \hat{a})$. The amplitude of terms well outside of this range is small. For this reason we say that the operator $\hat{a}^{\dagger} \hat{a} - \langle  \hat{a}^{\dagger} \hat{a} \rangle$ is of the order $\Delta ( \hat{a}^{\dagger} \hat{a})$. Rewriting $\hat{H}_I^Q$ in terms of $\hat{a}^{\dagger}\hat{a}$ (by the Holstein-Primakoff transformations), expanding the square roots (\ref{eq:end}) in $\hat{H}_I^Q$ around $\sqrt{\langle  \hat{a}^{\dagger} \hat{a}  \rangle}$ and keeping leading terms allows us to approximate

\begin{equation}  \hat{H}_I^Q \approx \lambda \sqrt{N_A} \left( \sqrt{\langle  \hat{a}^{\dagger} \hat{a}  \rangle} + \frac{\hat{a}^{\dagger}\hat{a}}{\sqrt{\langle  \hat{a}^{\dagger} \hat{a}  \rangle}}  \right) \hat{J}_B^x \approx \lambda \left( N_A |\zeta| + \frac{1}{|\zeta|} \hat{a}^{\dagger} \hat{a} \right)  \hat{J}_B^x ,  \end{equation} which is valid when 

\begin{equation} \lambda t \ll \frac{2\pi \sqrt{\langle  \hat{a}^{\dagger} \hat{a}  \rangle}}{N_B \sqrt{N_A}} \approx \frac{2\pi |\zeta|}{N_B} \quad \mbox{and} \quad \lambda t \ll \frac{2\pi \sqrt{N_A}}{ \langle \hat{a}^{\dagger} \hat{a} \rangle^{3/2} } \approx \frac{2\pi}{N_A |\zeta|^3}  . \label{eq:TCapprox1} \end{equation}

Transforming to the original basis, we find that the resultant state at time $t$ is equal to

\begin{equation}  \ket{\psi_m (t)} =   \hat{Q} e^{-it \hat{H}_I^Q} \hat{Q}^{-1}\ket{\psi_m (0)} \approx  \ket{\Psi_{m}(t)}_A \ket{D_m^{\phi} (t)}_B , \end{equation} a separable state of the composite system $AB$ where the state  

\begin{equation} \ket{\Psi_{m}(t)}_A \equiv  e^{-i t \lambda m N_A |\zeta| } \ket{\zeta e^{-it \lambda m / |\zeta|}}_{A} , \label{eq:psimt} \end{equation} which is a SCS of $A$ with a phase that is changing at a rate that is proportional to the value of $m$, and

\begin{equation}  \ket{D_m^{\phi} (t)}_B =  e^{-  it \lambda m (\hat{J}_B^z +c) / |\zeta| }\ket{D_m^{\phi}(0)}_B , \label{eq:Dmt} \end{equation} is a Dicke state in a rotating basis.
Thus, if the total initial state $\ket{\psi (0)}$ contains a superposition of $\ket{\psi_m (0)}$ with different values of $m$, the resultant state $\ket{\psi (t)}$ is, in general, an entangled state of the spin system $A$ and the ancillary spin system $B$.

For the simplest case of $N_B = 1$, $m$ takes values $\pm 1/2$ and we have

\begin{equation} \ket{\Psi_{\pm 1/2}(t)} = e^{\mp i t \lambda N_A |\zeta|/2 } \ket{\zeta e^{\mp it \lambda/ 2|\zeta|}}_{A},
\end{equation}
\begin{equation}
\ket{D_{\pm 1/2}^{\phi}(t)}_B =  \frac{1}{\sqrt{2}} \left( e^{-i\phi/2} e^{\mp i t \lambda / 2|\zeta|} \ket{\uparrow}_B \pm  e^{i\phi/2} \ket{\downarrow}_B \right) . \label{eq:JCapprox} \end{equation} This is the spin analogue of Gea-Banacloche's approximation for the Jaynes-Cummings model for strong initial coherent state and short times \cite{Gea-91}, as discussed in \cite{Shane13}. In the bosonic limit (making the transformations $\lambda \to {\lambda}/\sqrt{N_A}$ and $\zeta \to {\zeta}/\sqrt{N_A}$ and taking the $N_A \to \infty$ limit) we recover his results.

The initial separable state 

\begin{equation}   \ket{\psi (0)}  = \frac{1}{\sqrt{2}} \left( \ket{\psi_{+1/2}(0)} +  \ket{\psi_{-1/2}(0)} \right)  = \frac{1}{\sqrt{2}} \ket{\zeta}_{A} \left( \ket{D_{+1/2}^{\phi}(0)}_B + \ket{D_{-1/2}^{\phi}(0)}_B\right) \end{equation} evolves to the entangled state:

\begin{equation}   \ket{\psi (t)} \approx \mathcal{N} \left( \ket{\Psi_{+1/2}(t)}_A\ket{D_{+1/2}^{\phi}(t)}_B +  \ket{\Psi_{-1/2}(t)}_A \ket{D_{-1/2}^{\phi}(t)}_B \right) , \label{eq:JCcat} \end{equation} with macroscopic components $\ket{\Psi_{- 1/2}(t)}_A$ and $\ket{\Psi_{+ 1/2}(t)}_A$. The factor $\mathcal{N}$ is for normalisation since the two components $\ket{\Psi_{\pm 1/2}(t)}_A \ket{D_{\pm 1/2}^{\phi}(t)}_B$ are not , in general, orthogonal.

Although the approximation is valid only for short times in Eq.~(\ref{eq:TCapprox1}), numerics show that the qualitative features of Eq.~(\ref{eq:JCapprox}) persist for longer times. In particular, Eq.~(\ref{eq:JCapprox}) predicts that at time $\lambda t_r = 4 \pi |\zeta|$, the state of the spin system and the state of the qubit return to their respective initial states: $\ket{\Psi_{\pm} (t_r)}_A = \ket{\Psi_{\pm} (0)}_A $ and $\ket{D_{\pm 1/2}^{\phi}(t_r)}_B=\ket{D_{\pm 1/2}^{\phi}(0)}_B$. For the Jaynes-Cummings model, this $t_r$ is known as the revival time.

At a quarter of the revival time, the ancillary spin states is given by $\ket{D_{+ 1/2}^{\phi}(t_r/4)}_B = \ket{D_{-1/2}^{\phi}(t_r/4)}_B$ coincidently. In the context of the Jaynes-Cummings model, this is known as the ``attractor state''  of the qubit \cite{TPS09}. At this time, the state in Eq. (\ref{eq:JCcat}) becomes a separable state between $A$ and $B$ and the state in mode $A$ is a spin cat state:
\begin{equation}   
\ket{\psi (t_r/4)} \approx \left(\ket{\Psi_{+1/2}(t_r/4)}_A - \ket{\Psi_{-1/2}(t_r/4)}_A \right) \ket{D_{+ 1/2}^{\phi}(t_r/4)}_B. \end{equation}

For $N_B > 1$ we consider the initial separable state 

\begin{eqnarray}
\ket{\psi (0)} &=& \frac{1}{\sqrt{2}} \left( \ket{\psi_{+N_B/2}(0)} +  \ket{\psi_{-N_B/2}(0)} \right), \\
&=& \frac{1}{\sqrt{2}} \ket{\zeta}_{A} \left( \ket{D_{+N_B/2}^{\phi}(0)}_B + \ket{D_{-N_B/2}^{\phi}(0)}_B \right). \nonumber
\end{eqnarray}
This evolves to the entangled state:

\begin{equation}   \ket{\psi (t)} \approx \mathcal{N} \left( \ket{\Psi_{+N_B/2}(t)}_A\ket{D_{+N_B/2}^{\phi}(t)}_B +  \ket{\Psi_{-N_B/2}(t)}_A\ket{D_{-N_B/2}^{\phi}(t)}_B \right) , \label{eq:TCcat} \end{equation} with macroscopic components $\ket{\Psi_{- N_B/2}(t)}_A$ and $\ket{\Psi_{+ N_B/2}(t)}_A$. From Eq. (\ref{eq:psimt}) and Eq. (\ref{eq:Dmt}) we can see that at time $t = t_r / N_B$ the system revives to its initial state and at $t = t_r / 4N_B$ the big spin and the ancillary system are in a separable state with the big spin in a spin cat state:

\begin{equation}   
\ket{\psi (t_r/N_B 4)} \approx \left(\ket{\Psi_{+N_B/2}(t_r/4)}_A - \ket{\Psi_{-N_B/2}(t_r/4)}_A \right) \ket{D_{+ N_B/2}^{\phi}(t_r/4)}_B. \end{equation} We note that the revival time and spin cat state generation time are scaled by a factor of $1/N_B$ compared to $N_B = 1$, but that this speedup comes at the cost of having to prepare the ancillary system in a GHZ-like state. These results are consistent with the Tavis-Cummings model for large initial coherent state \cite{TPS09,Meu-06}. As for $N_B = 1$, although our approximation is strictly valid only for short times, it captures the qualitative features of the state at later times. 

The above results are not surprising, since we expect the $A$ system to look like a field mode in the bosonic limit. Outside this parameter regime, however, they dynamics of the system are not obvious. In \cite{Doo-14} it has been shown for the case $N_B=1$ that when $|\zeta|=1$ for the initial spin coherent state of $A$, the system evolves to spin squeezed states and to ``multiple cat states'', superpositions of two or more spin coherent states. Here we consider $|\zeta| = 1$ for the initial state of $A$, but with $N_B > 1$. In this case the Dicke state distribution is narrowly peaked around its average value $\langle \hat{a}^{\dagger}\hat{a} \rangle = \frac{N_A}{2}$ with $\Delta \hat{a}^{\dagger}\hat{a} = \sqrt{N_A}/2$. Alternatively we may say $\langle \hat{J}_A^z \rangle = 0$ with $\Delta \hat{J}_A^z =  \sqrt{N_A}/2$. The operator $\hat{J}_A^z$ appearing in $\hat{H}_I^Q$ is thus of the order $\Delta \hat{J}_A^z$. Expanding the square roots in $\hat{H}_I^Q$ around $\sqrt{\hat{J}_A^2} = \sqrt{\frac{N_A}{2}\left( \frac{N_A}{2} + 1 \right)} \approx \frac{N_A}{2} \gg 1$ and keeping leading terms allows us to approximate:

\begin{equation}  \hat{H}_I^Q \approx \lambda \left( 2\sqrt{\hat{J}_A^2} - \frac{(\hat{J}_A^z)^2 }{\sqrt{\hat{J}_A^2}} \right) \hat{J}_B^{x}  , \label{eq:Happrox2} \end{equation} valid when \begin{equation}  \lambda t \ll \frac{2\pi \sqrt{N_A}}{N_B}. \label{eq:timeapprox2} \end{equation} 

\begin{figure}[h]
\includegraphics[width=\columnwidth]{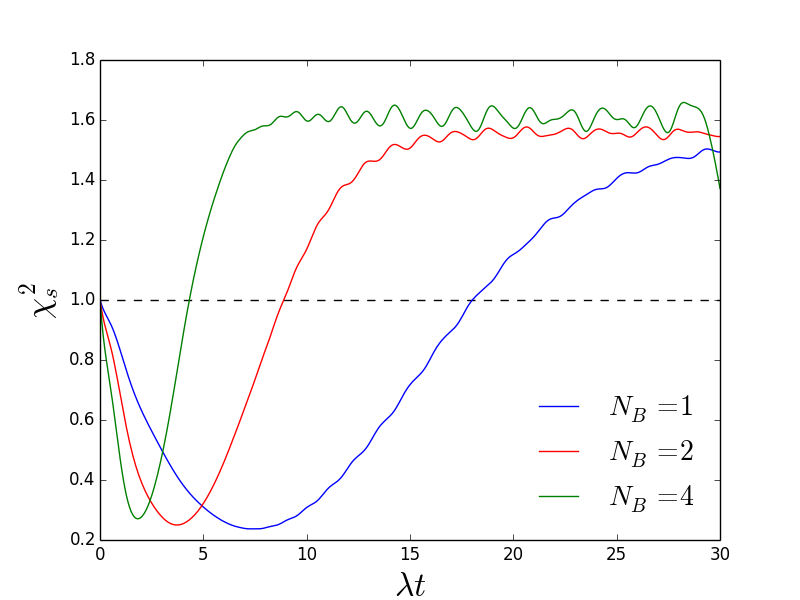}
\caption{The spin squeezing of $N_A = 80$ spins due to interaction with $N_B$ ancillary spins. For short times, when $\chi_s^2 < 1$, the state of $A$ is squeezed.}
\label{fig:squeeze}
\end{figure}

The second term in (\ref{eq:Happrox2}) is quadratic in $\hat{J}_A^z$ and is know as a \emph{finite Kerr} or \emph{one-axis twisting} term \cite{Kit-93,Ma-11}. It is well known that this term leads to spin squeezing of system $A$ with the most squeezing achieved at the short-time evolution. This can be quantified by the spin squeezing parameter proposed by Kitagawa and Ueda, $\chi_s^2 = 4\min ((\Delta \hat{J}_A^{\vec{n}_{\perp}})^2)/N_A$ where $\vec{n}$ is the mean spin direction of the state of the spin system, $\hat{J}_A^{\vec{n}_{\perp}}$ is the $J$-operator along an axis perpendicular to $\vec{n}$ and the minimisation is over all directions $\vec{n}_{\perp}$ perpendicular to $\vec{n}$ \cite{Kit-93,Ma-11}. In figure \ref{fig:squeeze} we plot $\chi_s^2$ as a function of time for $N_A = 80$ and initial state $\left[ \frac{1}{\sqrt{2}}\left( \ket{\downarrow} + \ket{\uparrow} \right) \right]^{\otimes N_A +N_B}$ of the combined $AB$ system. This is a spin coherent state of the $N_A + N_B$ spins with $\zeta = 1$ and we see spin squeezing of the state of $A$ at short times. For example, we observe the fact that the squeezing is $1/N_B$ faster in figure \ref{fig:squeeze} because the interaction strength in Hamiltonian \ref{eq:Happrox2} in enhanced by a factor of $N_B$ when the initial state is an eigenstate of $\hat{J}_B^x$.

\begin{figure}[]%[ht]
\centering
\subfigure{
   \includegraphics[width=45mm]{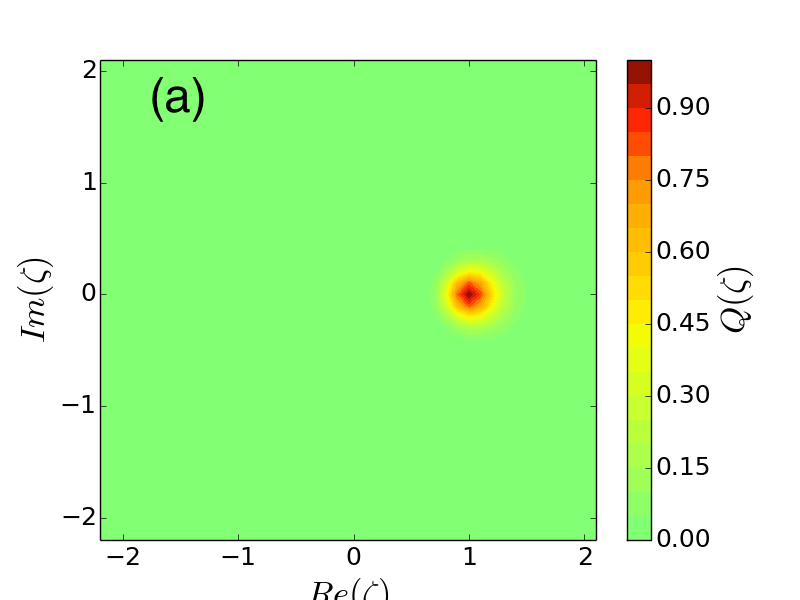}
   % \label{fig:SpWig5}
}
\subfigure{
	\includegraphics[width=45mm]{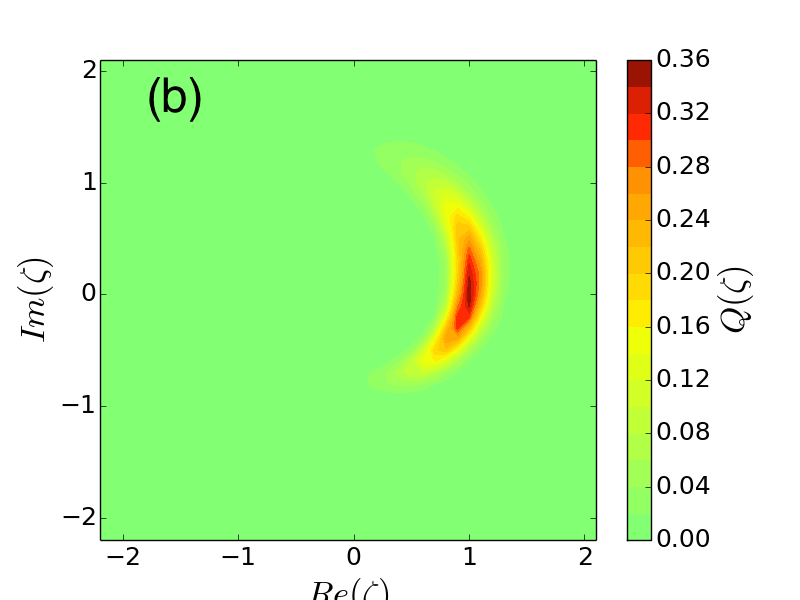}%{placeholder.jpg}
    %\label{fig:SpWig10}
}
\subfigure{
	\includegraphics[width=45mm]{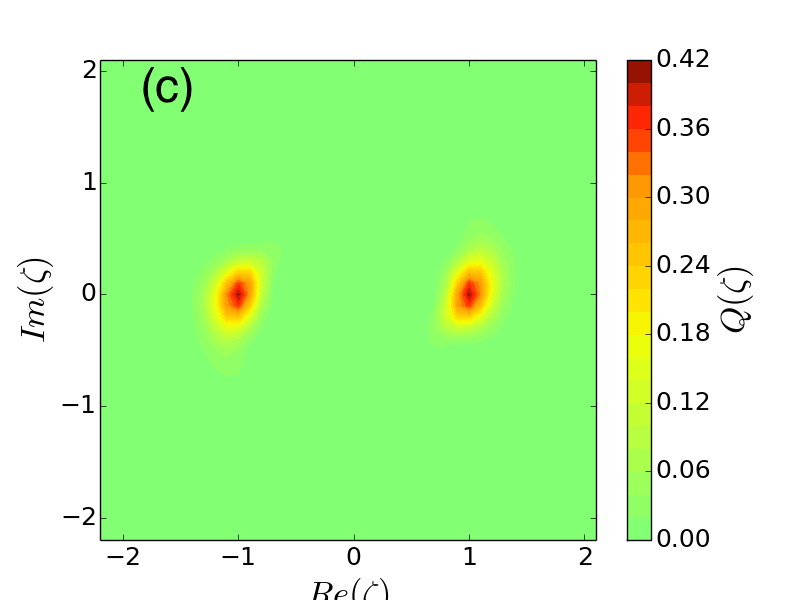}%{placeholder.jpg}
    %\label{fig:SpWig10}
}
\subfigure{
	\includegraphics[width=45mm]{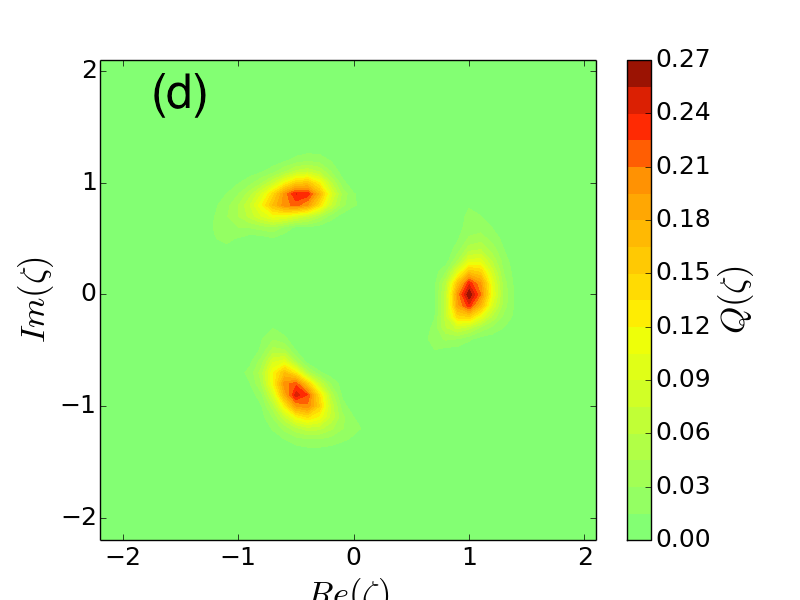}%{placeholder.jpg}
    %\label{fig:SpWig10}
}
\subfigure{
	\includegraphics[width=45mm]{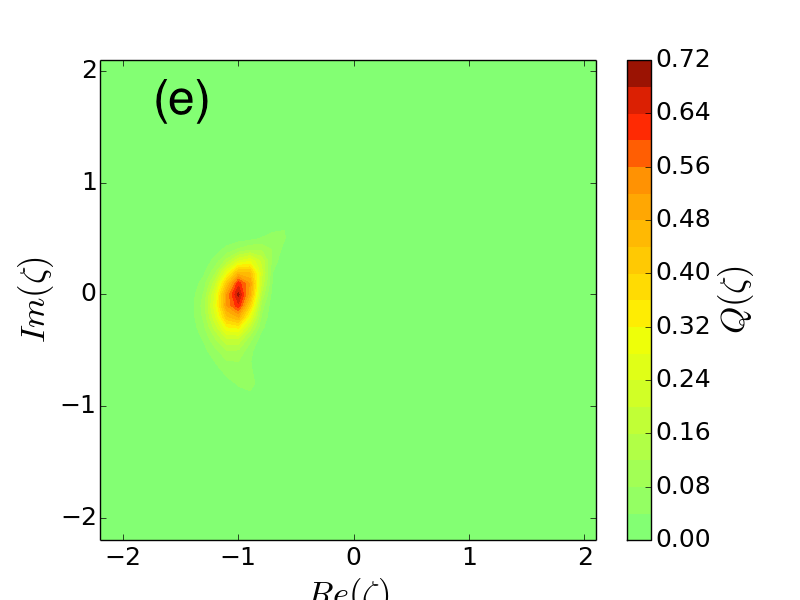}%{placeholder.jpg}
    %\label{fig:SpWig10}
}
\subfigure{
	\includegraphics[width=45mm]{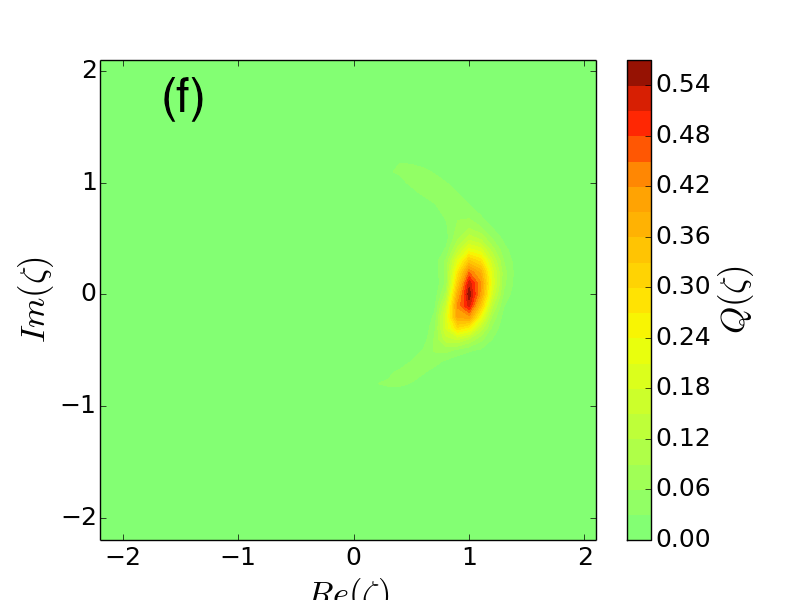}%{placeholder.jpg}
    %\label{fig:SpWig10}
}
\caption[]{$Q$-functions for the combined $N_A + N_B$ spin system with $N_A = 80$ and $N_B=2$ ($\zeta=1$). (a) at $t=0$, (b): a spin squeezed state at $t=T/40$, (c): a GHZ state at $t=T/4$, (d): a three component spin cat state at $t=T/3$, (e) the ``anti-revival'' of the initial state at $t=T/2$, and (f): the revival at $t=T$. 
}
\label{fig:Qfunctions}
\end{figure}

\begin{figure}[h]
\includegraphics[width=\columnwidth]{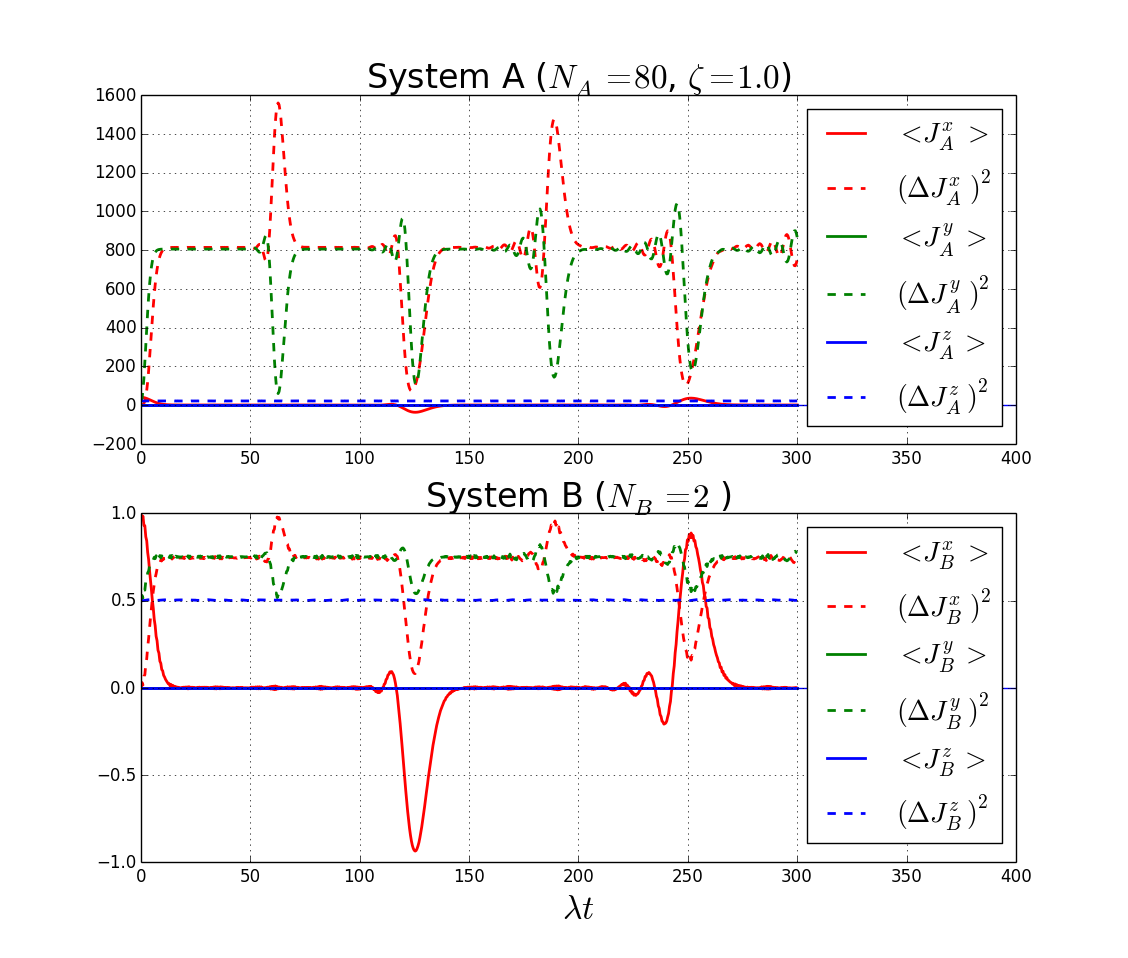}
\caption{Expectation values and standard deviations in time evolution for $N_A = 80$ and $N_B = 2$ ($\zeta=1$). At the time {\bf\blue $\lambda t = \lambda T/4 \approx 63$}, $\Delta J^x_A$ and $\Delta J^x_B$ are close to their maximum, which indicates that two systems are maximally entangled.  }
\label{fig:ExpVarAB}
\end{figure}

Figure \ref{fig:Qfunctions} shows $Q$-functions for the combined $N_A+N_B$ spin state at various times. The $Q$-function for a state $\ket{\psi}$ of the composite $AB$ system is defined as $Q(\zeta) = |\langle \psi | \zeta \rangle_{AB}|$ where $\ket{\zeta}_{AB}$ is a spin coherent state of the $N_A + N_B$ spins. In figure \ref{fig:Qfunctions}(a) we show the $Q$ function for the initial state $\left[ \frac{1}{\sqrt{2}}\left( \ket{\downarrow} + \ket{\uparrow} \right) \right]^{\otimes N_A +N_B}$ of the system. After a short time, the system has evolved to a spin squeezed state, as illustrated in figure \ref{fig:Qfunctions}(b) and as predicted by Eq. (\ref{eq:Happrox2}). Figure \ref{fig:Qfunctions}(f) shows that after a period $T= 2\pi N_A / \lambda N_B$ the system has returned (approximately) to its initial spin coherent state. At $t=T/4$ [figure \ref{fig:Qfunctions}(c)] the system is in a GHZ state, a superposition of the spin coherent states $\left[ \frac{1}{\sqrt{2}}\left( \ket{\downarrow} + \ket{\uparrow} \right) \right]^{\otimes N_A +N_B}$ and $\left[ \frac{1}{\sqrt{2}}\left( \ket{\downarrow} - \ket{\uparrow} \right) \right]^{\otimes N_A +N_B}$, while at $t=T/3$ [figure \ref{fig:Qfunctions}(d)] it is a ``multiple-cat'' state, a superposition of three spin coherent states. The $Q$-function at half the revival time [figure \ref{fig:Qfunctions}(e)] shows the spin coherent state $\left[ \frac{1}{\sqrt{2}}\left( \ket{\downarrow} - \ket{\uparrow} \right) \right]^{\otimes N_A +N_B}$, an ``anti-revival'' of the initial state since each spin has been flipped relative to $\ket{\psi (0)}_{AB}$. In figure \ref{fig:ExpVarAB} we plot expectation values and variances for $J$-operators of $A$ and $B$ for the same initial state and for $N_A = 80$, $N_B = 2$. We see the revival of each of the plotted quantities at $\lambda t = \lambda T\approx 250$. At $t = T/4$ the variance of the system in the operator $\hat{J}_A^x$ is close to its maximum value $N_A^2 / 4 = 1600$. Similarly, at this time the variance in $\hat{J}_B^x$ is close to its maximum of $N_B^2/4 = 1$. This indicates that the system is in a GHZ state of the combined $N_A + N_B$ spins. 

\subsection{Ansatz for the dynamics}
We find by ansatz and by comparison with numerics that the Hamiltonian

\begin{equation}   \hat{H}_I^Q   =  \lambda \left( N_A - \frac{2(\hat{J}_A^z)^2 + 4c \hat{J}_A^z}{N_A} \right) \hat{J}_B^{x}  , \label{eq:Happrox3}   \end{equation} predicts many of the gross features of the exact dynamics of the system for initial state $\ket{\psi (0)}_{AB} = \left[ \frac{1}{\sqrt{2}}\left( \ket{\downarrow} + \ket{\uparrow} \right) \right]^{\otimes N_A +N_B}$.

Assuming for simplicity that $N_A$ and $N_B$ are even, we expand the initial state in the Dicke basis $\ket{n}_A$ and evolve by the Hamiltonian in Eq.~(\ref{eq:Happrox3}) to get: 
\begin{eqnarray} \ket{\psi (t)}_{AB} &=& \hat{Q}e^{-it\hat{H}_I^Q} \hat{Q}^{-1}\ket{\psi (0)}_{AB} \label{eq:blah} \\ &\approx&  \hat{Q}e^{-i t\lambda {N_A N_B \over 2}} \sum_{n=0}^{N_A} C_n \exp \left[ \frac{i t\lambda N_B \left(n- \frac{N_A}{2}\right)^2 }{N_A} \right] \ket{n}_{A}  \left[ { \ket{\downarrow} + \ket{\uparrow} \over \sqrt{2}} \right]^{\otimes N_B}. \nonumber \end{eqnarray} Since the time dependent exponentials in Eq.~(\ref{eq:blah}) are periodic in time with period $T = 2 \pi N_A / \lambda N_B$, the state in Eq.~(\ref{eq:blah}) itself is periodic in time with the same period (if $N_A$ is odd the period is $4T$). This implies that \begin{equation} \ket{\psi (T)}_{AB} = \hat{Q} e^{-i T\hat{H}_I^Q} \hat{Q}^{-1}\ket{\psi (0)}_{AB} \approx  \ket{\psi (0)}_{AB} , \label{eq:blah2} \end{equation} the system revives to its initial state. 

More generally, we can find a useful expression for the state of the system at times $t=pT/q$ that are rational fractions of the revival time ($p$ and $q$ are coprime integers) \cite{Ave-89}. At such times the exponential $\exp\left[ \frac{2 i p \pi}{q} \left(n- \frac{N_A}{2}\right)^2 \right]$ in Eq.~(\ref{eq:blah}) is a periodic function of the discrete variable $n$ with period $q$. This means that we can express the exponential in terms of its discrete Fourier transform $\mathcal{F}_l$: 
\begin{equation}  \exp\left[ \frac{2 i p \pi}{q} \left(n- \frac{N_A}{2}\right)^2 \right] = \frac{1}{\sqrt{q}} \sum_{l=0}^{q-1} \mathcal{F}_l \, \exp\left[ \frac{-2\pi i l n}{q} \right], \label{eq:dftinv} \end{equation}
where the discrete Fourier transform is given by
\begin{equation}  \mathcal{F}_l = \frac{1}{\sqrt{q}} \sum_{n=0}^{q-1}  \exp\left[ \frac{2 i p \pi}{q} \left(n- \frac{N_A}{2}\right)^2 \right] \, \exp\left[ \frac{2\pi i l n}{q} \right] . \label{eq:dft} \end{equation} %is known as a \emph{generalised quadratic Gauss sum} \cite{xx}. 

Substituting Eq.~(\ref{eq:dftinv}) into Eq.~(\ref{eq:blah}) allows us to write the state of the system at $t=pT/q$ as a superposition of spin coherent states of the combined $N_A + N_B$ spin system: 

\begin{equation} 
\ket{\psi (pT/q)}_{AB} = e^{-i pT\lambda N_B N_A / 2q} \frac{1}{\sqrt{q}} \sum_{l=0}^{q-1} \mathcal{F}_l \, e^{i\pi l N_B /q} \left[ { \ket{\downarrow} + e^{\frac{-2\pi i l}{q}} \ket{\uparrow} \over \sqrt{2}} \right]^{\otimes N_A + N_B} . \label{eq:OATpsi}\end{equation} 

The sum in Eq. (\ref{eq:dft}) is known as a \emph{generalised Guass sum} \cite{Ber-81} and can be calculated for various values of $p$ and $q$. Suppose, for example, that $p=1$ and $q=4k$ for some integer $k$. This means that we are looking at the state at times $t \in \left\{ \frac{T}{4}, \frac{T}{8}, \frac{T}{12},...   \right\}$. In this case (ignoring global phase factors) we have

\begin{equation} \ket{\psi (T/4k)}_{AB} \propto \frac{(1+i)}{\sqrt{4k}} \sum_{l = 0}^{2k-1}  f_l \left[ { \ket{\downarrow} + e^{\frac{-\pi i l}{k}} \ket{\uparrow} \over \sqrt{2}} \right]^{\otimes N_A + N_B}, \end{equation}
for $f_l =  \exp\left[ \frac{i\pi (2l-N_A)^2}{8k} \right] \exp\left[ \frac{i\pi l N_B }{2k} \right]$.

This is an equally weighted superposition of $2k = q/2$ spin coherent states distributed uniformly in phase. This is consistent with figure \ref{fig:Qfunctions}(c), which shows a superposition of two spin coherent states at $t=T/4$. Similar expressions can be derived for other values of $p$ and $q$ and for $N_A$ or $N_B$ odd. For each case that was checked against numerics with initial state $\ket{\psi (0)}_{AB}$, the Hamiltonian in Eq.~(\ref{eq:Happrox3}) predicts the correct revival time, anti-revival time, and the correct time for the generation of a superpositions of two spin coherent states [as at $t=T/4$ in figure \ref{fig:Qfunctions}(c)].

\section{Implementation for entangled SCSs}
\label{Sec_4}

Non-classical entangled states of spins in atomic ensembles have been successfully demonstrated in hot Rubidium gases \cite{Polzik10,Polzik99,Polzik14} and NV centres \cite{Bill11} for qubit-spin interaction in the field of quantum information. In particular, cold-atom schemes with BECs is one of the strongest candidates for a prototype of a scalable quantum information processor and is naturally fit for CV SCS schemes \cite{Wineland,Byrnes12}.

\subsection{Entangling SCSs through spin BS interaction}
If we prepare the initial state as a non-classical superposed SCS in one spatial mode $A$ and a SCS in the other mode $B$, the key implementation element for entangling operation is a spin BS interaction $\hat{H}_I (N_A, N_B)$ given in Eq.~(\ref{eq:Hint}), which is also used for generating a superposed SCS. In the optical regime it is well known that such an interaction can create interesting path-entangled states for suitable non-classical input states. An example is entangled coherent states (ECSs) \cite{ECS1,ECS4,ECS5}, which have been demonstrated to be a useful resource for quantum enhanced metrology \cite{ECS2,ECS3}. These are created by the mixing of a Schr\"odinger cat-type state and a coherent state on a BS and by analogy to the optical case, similar states may be created in spin systems.

We here consider how to create entangled states in BECs. Since the phenomenon of BECs in dilute boson gases was proposed theoretically in the early 1920s, it has been actively demonstrated for the last two decades experimentally. When the temperature or volume of the gas reaches a critical point, the condensate phenomenon is observed in the number of particles in the gas at very low temperature. In the simplified model of trapped BEC atoms by ac-Stark effect (e.g., in two counter-propagating trapping lasers), the BEC atoms are occupied in two hyperfine ground states ($\ket{g_1}$ and $\ket{g_2}$), which can be coupled with an excited state $\ket{e}$ \cite{Jannis-Peter}. Then, initialised BEC atoms in mode $A$ is given by
\begin{equation}
\ket{\zeta}_A
= {1\over \sqrt{1+|\zeta|^2}} \left( \ket{g_1}_A + \zeta \ket{g_2}_A \right)^N.
\label{BEC_01}
\end{equation}

Suppose that two BECs are trapped in a cavity (e.g., a superposed SCS in mode $A$ and a typical SCS in mode $B$). Circular polarised light coupling between $\ket{e}$ and $\ket{g_1}$ (or $\ket{g_2}$) perform an adiabatic passage between  $\ket{g_1}$ and $\ket{g_2}$, and then, a cavity field can mediate two BECs to obtain an effective spin Hamiltonian \cite{Byrnes12}
\begin{eqnarray}
&& \hat{H}_{BEC} \approx \Omega \left( J^{+}_{A} J^{-}_{B} + J^{-}_{A} J^{+}_{B} \right),
\label{BEC_05}
\end{eqnarray} 
after adiabatic elimination of the high-energy level $\ket{e}$ in the cavity ($\Omega$ is the effective Rabi frequency). 

Note that the same spin interaction is used for creating superposed spin states between a SCS in BECs and ancillary spins. For example, a well-focused laser beam creates a micro-trap (e.g., an optical tweezer trap) which can contain a single atom in an optical potential \cite{optical_tweezer01}. By moving the trapping potential, the ancillary atom can be shifted into the interaction region of a cavity with the prepared SCS and the same effective Hamiltonian is given for building a non-classical superposed SCS in the cavity. Alternative methods of Schr\"odinger cat generation have been recently proposed using polarised light in trapped spins \cite{New_super_BEC13}. Particularly, one of the superposed spin states can be built as a GHZ-type spin state during the time evolution and can be used for an initial state, which can be converted into a macroscopic optical Schr\"odinger cat state using the Tavis-Cummings Hamiltonian \cite{TPS09}.

In addition, one might use the entanglement between a SCS and a qubit to entangle two SCSs. If the qubit is represented by a polarised light interacting with a hot atomic ensemble \cite{New_super_BEC13,Polzik10,Polzik99,Polzik14} or superconducting flux qubit connected with NV-centres \cite{NV-Lukin10,Bill11}, we could generate two entangled states given by a pair of $(\ket{\zeta} \ket{0} + \ket{-\zeta} \ket{1})/\sqrt{2}$. Then, if a Bell state measurement is performed in two ancillary qubits, the outcome state in two SCSs is equivalent to the ECS.

\section{Summary and remarks}
\label{Sec_5}
In summary, we have investigated the generation of non-classical superposed spin states using ancillary spins. Our focus has been on the scheme of generating two or multiply superposed SCSs for $N_B = 2$ and propose a generalised Hamiltonian for $N_B > 2$ that explains the gross features of the model. We also addressed the implementation of entangled SCSs (e.g., in BECs or NV-centres) in the light of optical techniques.

For practical implementation, the decoherence mechanism needs to be understood for different physical systems. For example, the general decoherence effects in BECs could be Markovian dephasing on each atoms and particle losses \cite{Byrnes12,byrnes2013fractality} while NV-centres and hot atomic gases are confined without losing spins. In optics the major decoherence model is based on particle loss and this can be simulated by additional BSs \cite{ECS2,ECS3}. The analogue of photon loss in the spin system is understood in terms of spin flips since the ``vacuum'' corresponds to all the spins in state $\ket{\downarrow}$. The speedup of preparation could be degraded under strong decoherence in order to create spin cat states. In addition, robust quantum memories in ensembles is one of the key ingredients for practical quantum information processing \cite{Lukin_RMP} and has been actively investigated in spin systems \cite{Molmer_PRL13,Molmer_PRA13,Molmer_PRA10}. These approaches could be applied in CV spin entangled states in order to develop practical CV quantum information processing in the future.

\section{Acknowledgements}
We thank Paul Knott, Francis McCrossan and Anthony Hayes for useful discussion at an early stage of this research.

\end{document}